\documentclass[pra,twocolumn,tightenlines,showpacs,showkeys]{revtex4}

\usepackage{graphicx}
\usepackage{bbm}
\usepackage{amsfonts}
\usepackage{theorem}
\usepackage{amssymb}
\usepackage{amsmath}
\usepackage{bbm}

\newcommand{\beq}{\begin{eqnarray}}
\newcommand{\eeq}{\end{eqnarray}}

\begin{document}
\title{A simplex of bound entangled multipartite qubit states}
\author{B.C.~Hiesmayr},
\author{F.~Hipp},
\author{M.~Huber},
\author{Ph.~Krammer},
\author{Ch.~Spengler}
\address{Fakult\"at f\"ur Physik, Universit\"at Wien, Boltzmanngasse 5, A-1090 Vienna,
Austria}

\begin{abstract}
We construct a simplex for multipartite qubit states of even number
$n$ of qubits, which has the same geometry concerning separability,
mixedness, kind of entanglement, amount of entanglement and
nonlocality as the bipartite qubit states. We derive the
entanglement of the class of states which can be described by only
three real parameters with the help of a multipartite measure for
all discrete systems. We prove that the bounds on this measure are
optimal for the whole class of states and that it reveals that the
states possess only $n$--partite entanglement and not e.g. bipartite
entanglement. We then show that this $n$--partite entanglement can
be increased by stochastic local operations and classical
communication to the purest maximal entangled states. However, pure
$n$--partite entanglement cannot be distilled, consequently all
entangled states in the simplex are $n$--partite bound entangled. We
study also Bell inequalities and find the same geometry as for
bipartite qubits. Moreover, we show how the (hidden) nonlocality for
all $n$--partite bound entangled states can be revealed.
\end{abstract}


\maketitle

\section{Introduction}

Entanglement is at the heart of the quantum theory. It is the source
of several new applications as quantum cryptography or a possible
quantum computer. In recent years by studying higher dimensional
quantum systems and/or multipartite systems one realizes that
different aspects of the entanglement feature arise. They may have
new applications such as multiparty cryptography.

In this paper we contribute to the classification of entanglement in
a twofold way, i.e. which kind of entanglement a certain class of
multipartite qubit states possesses by using the multipartite
measure proposed in Ref.~\cite{HH2} and whether this kind of
entanglement can be distilled. Our results suggest that one can
distinguish for multipartite systems between different
possibilities.

The class of states we analyze are a generalization of the class of
states which form the well--known simplex for bipartite qubits
 (Sec.~\ref{simplexqubits}), i.e. all
locally maximally mixed states \cite{BNT02,Horodecki}. We make an
obvious generalization and find for states composed of an even
number of qubits $n$ an analogous simplex, i.e. this class of states
shows the same geometry concerning positivity, mixedness,
separability and entanglement (Sec.~\ref{simplex}). Further the used
multipartite measure \cite{HH2} reveals that the kind of
entanglement possessed is only a $n$--partite entanglement where $n$
is the number of qubits involved. The vertex states of the simplex
are represented in the bipartite case by the well known Bell states,
for $n>2$ they are equivalent to the generalized smolin states
proposed by
Ref.~\cite{Smolin,HorodeckigeneralizedSmolin2,HorodeckigeneralizedSmolin,Bandyopadhyay04}.

Then we discuss the distillability of the entangled states and find
states for which the $n$--partite entanglement can be increased by a
protocol only based on copy states and stochastic local operations
and classical communications (LOCC). We show that the state is not
distillable for any subset of parties and hence bound entangled,
however, the $n$--partite entanglement can be enhanced to reach the
maximal possible purity and  $n$--partite entanglement within the
class of states under investigation, i.e. the vertex states. For a
subset of these states it has been shown that they allow for quantum
information concentration ,e.g.
Ref.~\cite{HorodeckigeneralizedSmolin2,muraovedral}, so we suggest
that it might still be advantageous to enhance the $n$--partite
bound entangled states for some applications.

Last but not least, in Sec.~\ref{chshsimplex} we address to the
question which of the simplex states violate the generalized Bell
inequality which was shown to be optimal in this case and draw its
geometrical picture, Fig.~\ref{chshfig}.

\section{The simplex for bipartite qubits}\label{simplexqubits}

A single qubit state $\omega$ lives in a two dimensional Hilbert
space, i.e. ${\cal H}\equiv\mathbb{C}^2$, and any state can be
decomposed into the well known Pauli matrices $\sigma_i$
\begin{equation*} \label{qubitpauli}
\omega \;=\; \frac{1}{2} \left( \mathbbm{1}_2 + n_i\, \sigma_i
\right)
\end{equation*}
with the Bloch vector components $\vec{n} \in \mathbbm{R}^3$ and
$\sum_{i=1}^3 n_i^2 = \left| \vec{n} \right|^2 \leq 1$. For
$\left|\vec{n}\right|^2 < 1$ the state is mixed (corresponding to
Tr$\,\omega^2 < 1$) whereas for $\left|\vec{n}\right|^2 = 1$ the
state is pure (Tr$\,\omega^2 = 1$).

\begin{figure}
\center{\includegraphics[height=200pt,
keepaspectratio=true]{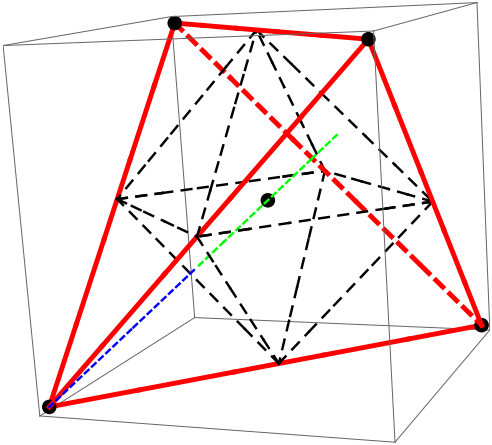}} \caption{(Color online) Here
the geometry of the state space of even number of qubits is
visualized. Each state is represent by a triple of three real
numbers, $\vec{c}$, Eq.~(\ref{defstates}). The four black dots in
the vertices of the cube represent four orthogonal ``vertex''
states. In the case of two qubits these are the four maximally
entangled Bell states $\psi^\pm,\phi^\pm$ and for higher $n$ they
are equally mixtures of $2^n/4$ $GHZ$--states. The positivity
condition forms a tetrahedron (red) with the four ``vertex'' states
and the totally mixed state in the origin (black dot in the middle).
All separable states are represented by points inside and at the
surface of the octahedron (dashed object). The dashed line
represents for $n=2$ the Werner states and for $n>2$ the generalized
Smolin states (becoming separable when blue changes into
green).}\label{qubitstetrahedron}
\end{figure}

The density matrix of $2$--qubits $\rho$ on ${\mathbb C}^2 \otimes
{\mathbb C}^2$ is usually obtained by calculating its elements in
the standard product basis, i.e.
$|00\rangle,|01\rangle,|10\rangle,|11\rangle$. Alternatively, we can
write any $2$--qubit density matrix in a basis of $4 \times 4$
matrices, the tensor products of the identity matrix $\mathbbm{1}_2$
and the Pauli matrices,
$$
\rho= \frac{1}{4} \left( \mathbbm{1}_2 \otimes \mathbbm{1}_2 +
    a_i\,\sigma_i \otimes \mathbbm{1}_2 + b_i\,\mathbbm{1}_2 \otimes \sigma_i +
    c_{ij}\,\sigma_i \otimes \sigma_j \right)
$$
with $ a_i,b_i,c_{ij} \in \mathbbm{R}$. The parameters $a_i, b_i$
are called \textit{local} parameters as they determine the
statistics of the reduced matrices, i.e. of Alice's or Bob's system.
In order to obtain a geometrical picture one considers in the
following only states where the local parameters are zero
($\vec{a}=\vec{b}=\vec{0}$), i.e., the set of all locally maximally
mixed states, $Tr_A(\rho)=Tr_B(\rho)=\frac{1}{2}\mathbbm{1}_2$ (see
also Ref.~\cite{BNT02,Horodecki}).

A state is called separable if and only if it can be written in the
form $\sum_i p_i\, \rho_i^A\otimes \rho_i^B$ with $p_i\geq 0,\sum
p_i=1$, otherwise it is entangled. As the property of separability
does not change under local unitary transformation and classical
communication (LOCC) the states under consideration can be written
in the form \cite{BNT02}
\begin{equation*}
\rho \;=\; \frac{1}{4}\left(\mathbbm{1}_2 \otimes
\mathbbm{1}_2+c_i\,\sigma_i \otimes \sigma_i \right)\, ,
\end{equation*}
where the $c_i$ are three real parameters and can be considered as a
vector  $\vec{c}$ in Euclidean space. Differently stated, for any
locally maximally mixed state $\rho$ the action of two arbitrary
unitary transformations $U_1\otimes U_2$ can via the homomorphism of
the groups $SU(2)$ and $SO(3)$ be related to unique rotations
$O_1\otimes O_2$. Thus the correlation matrix $c_{ij}\,
\sigma_i\otimes\sigma_j$ can be chosen such that the matrix $c_{ij}$
gets via singular value decomposition diagonal. Therefore, three
real numbers combined to a vector $\vec{c}$ can be taken as an
representative of the state itself.

In Fig.~\ref{qubitstetrahedron} we draw the $3$--dimensional
picture, where each point $\vec{c}$ corresponds to a locally
maximally mixed state $\rho$. The origin $\vec{c}=\vec{0}$
corresponds to the totally mixed state, i.e. $\frac{1}{4}
\mathbbm{1}_2 \otimes \mathbbm{1}_2$. The only pure states in the
picture are given by $|\vec{c}|^2=3$ and represent the four
maximally entangled Bell states
$|\psi^\pm\rangle=\frac{1}{\sqrt{2}}\{|01\rangle\pm|10\rangle\},|\phi^\pm\rangle=\frac{1}{\sqrt{2}}
\{|00\rangle\pm|11\rangle\}\;,$ which are located in vertices of the
cube. The planes spanned by these four points are equivalent to the
positivity criterion of the state $\rho$. Therefore, all points
inside the tetrahedron represent the state space.

It is well known that density matrices which have at least one
negative eigenvalue after partial transpose ($PT$)
are entangled. The inversion of the argument
is only true for systems with $2\otimes2$ and $2\otimes3$ degrees of
freedom. $PT$ corresponds to a reflection, i.e. $c_2\rightarrow-c_2$
with all other components unchanged. Thus all points inside and at
the surface of the octahedron represent all separable states in the
set. Of course, one can always make the transformation
$\vec{c}\longrightarrow-\vec{c}$, thus one obtains a mirrored
tetrahedron, spanned by the four other vertices of the cube.
Clearly, the intersection of the these two tetrahedrons contain all
states which have positive eigenvalues after the action of $PT$.

In Ref.~\cite{BHN3,BHN2,BHN1} a generalization to higher dimensional
bipartite states is considered and a so called magic simplex for
qudits is obtained. Here the class of all locally maximally mixed
states have to be reduced in order to obtain this generalized
simplex. Already for bipartite qutrits many new symmetries arise and
regions of bound entanglement can be found (see also
Refs.~\cite{BertlmannKrammer4,BertlmannKrammer3,BertlmannKrammer2,BertlmannKrammer,Derkacz}).

We also want to generalize the simplex of bipartite qubits, however,
in our case we increase the number of qubits.

\section{A simplex for $n$--partite qubit states}\label{simplex}

Assume we have $n$ qubits. Then a generalization can be written as
\beq\label{defstates} \rho&=&\frac{1}{2^n}\left(\mathbbm{1}+\sum
c_i\;
\sigma_i\otimes\sigma_i\otimes\cdots\otimes\sigma_i\right)\nonumber\\
&:=&\frac{1}{2^n}\left(\mathbbm{1}+\sum c_i\; \sigma_i^{\otimes
n}\right) \eeq  Obviously, for this generalization we follow the
strategy to set the local parameters of all subsystems $j$,
$Tr_{1,2,\dots,j-1,j+1,\dots, n}(\rho)$, to zero, as well as the
parameters shared by two parties $j,k$,
$Tr_{1,2,\dots,j-1,j+1,\dots,k-1,k+1,\dots,n}(\rho)$, zero and so on
until $n-1$ zero.

Again the state can be represented by a three dimensional vector
$\vec{c}$. For $n=3$ the positivity condition $\rho \geq 0$ requires
\beq|\vec{c}\,|^2\leq 1\;.\eeq This turns out to be the case for all
odd numbers of qubits involved.

For even numbers of qubits the positivity condition $\rho \geq 0$
requires that the vector is within the following four
planes~\footnote{Note that for $n=4,8,\dots$ the mirrored
tetrahedron ($\vec{c}\longrightarrow-\vec{c}$) is obtained.}:
\beq\label{positivity}&&1+\vec{c}\cdot\vec{n}^{\,(i)}\;\geq\;0
\qquad\\
&&\textrm{with}\quad\vec{n}^{\,(i)}\;=\;\left(\begin{array}{c}-1\\+1\\+1\\\end{array}\right),
\left(\begin{array}{c}+1\\-1\\+1\\\end{array}\right),
\left(\begin{array}{c}+1\\+1\\-1\\\end{array}\right),
\left(\begin{array}{c}-1\\-1\\-1\\\end{array}\right)\nonumber\eeq
These conditions are exactly the same ones as for the two qubit case
$n=2$, i.e. the four above planes form the magic tetrahedron.

The purity $Tr(\rho^2)$ gives $\frac{1}{2^n}(1+|\vec{c}|^2)$, thus
the states with $|\vec{c}|^2=3$ are the purest states of the class
of states under investigation and are located in the vertices of the
tetrahedron. Note with increasing $n$ the percentage of purity
decreases, i.e. only for $n=2$ the vertices present pure states.
Further analysis of these vertex states follows later.

Now we want to investigate if the separability condition also for
$n>2$ corresponds to the octahedron. The partial transpose of one
qubit ($PT_{\textrm{one qubit}}$) changes the sign in front of the
$\sigma_2^{\otimes n}$ matrix, i.e. the $y$--component of the vector
$\vec{c}$ changes sign. Therefore the states under investigation are
entangled by the necessary but not sufficient (one qubit) Peres
criterion \beq\label{PTsingle}
\textrm{n=3,5,\dots:}&&|\vec{c}\,|^2\leq 1\nonumber\\
\textrm{n=2,4,\dots:}&&1-\vec{c}\cdot\vec{n}^{\,(i)}\;\leq\;0\;.\eeq 
Taking the partial transpose of two, four, \dots qubits changes two,
four,\dots times the sign and consequently one obtains the
positivity criterion (\ref{positivity}). Taking the partial
transpose of odd qubits is equivalent to $PT_{\textrm{one qubit}}$.

For even number of qubits the above Peres criterion implies a
mirrored tetrahedron, analogously to the bipartite case, however, we
do not know if the intersection, the octahedron, contains only
separable states. For odd numbers of qubits the situation is
different and we will not investigate it further.


Now two questions arise, firstly, are all states represented by the
octahedron separable and, secondly, what kind of entanglement does
this class of states possess?

Let us tackle the second question first. To analyze our generalized
states $\rho$ further we use the multipartite entanglement measure
for all discrete systems introduced by Ref.~\cite{HH2}. The main
idea is that the information content of any $n$--partite quantum
system of arbitrary dimension can be separated in the following
form:
\begin{eqnarray}\label{voll}
\underbrace{I(\rho)+R(\rho)}_{\textrm{single
property}}+\underbrace{E(\rho)}_{\textrm{entanglement}}\;=\;n
\end{eqnarray}
where
\begin{eqnarray}
I(\rho)&:=&\sum_{s=1}^n\underbrace{{\cal
S}_s^2(\rho)}_{\textrm{single property of subsystem s}}
\end{eqnarray}
contains all locally obtainable information (i.e. obtainable
information a party can measure on its particle) and $E(\rho)$
contains all information encoded in entanglement and $R(\rho)$ is
the complementing missing information, it is due to a classical lack
of knowledge about the quantum state. The total amount of
entanglement $E(\rho)$ can be separated into $m$--flip concurrences
by rewriting the linear entropy of all subsystems in an operator
sum, thus one obtains
\begin{eqnarray}
E(\rho)&:=&\underbrace{\textbf{C}_{(2)}^2(\rho)}_{\textrm{two flip
concurrence}}+\underbrace{\textbf{C}_{(3)}^2(\rho)}_{\textrm{three
flip concurrence}}+\;\quad(\dots)\nonumber\\&&+\underbrace{\textbf{
C}_{(n)}^2(\rho)}_{\textrm{n-flip concurrence}}\,.
\end{eqnarray}
These $m$--flip concurrences are useful for two reason: firstly, one
can obtain bounds on the operators and thus handle mixed states and
secondly, the authors of Ref.~\cite{HH2} showed (for three qubits)
that the $m$--flip concurrences can be reordered such that they give
the $m$--partite entanglement, which in addition coincides with the
$m$--separability \cite{Horodeckmseparability}.

Here we extend their result for the states under investigation. Due
to high symmetry of the class of states under investigation the
bounds of the $m$--partite entanglement can be computed and herewith
we can reveal the following substructure of total entanglement
$E(\rho)$
\begin{eqnarray}
E(\rho)&=&\underbrace{E_{(2)}(\rho)}_{\textrm{bipartite
entanglement}}+\underbrace{E_{(3)}(\rho)}_{\textrm{tripartite
entanglement}}\nonumber\\
&&+\dots+\underbrace{E_{(n)}(\rho)}_{\textrm{$n$--partite
entanglement}}
\end{eqnarray}
with the sub--substructure
\begin{eqnarray}
E_{(2)}(\rho)&=&E_{(12)}(\rho)+E_{(13)}(\rho)+\dots+E_{(1n)}(\rho)\nonumber\\
\lefteqn{+E_{(23)}(\rho)+\dots+E_{(2n)}(\rho)+\dots +E_{(n-1,n)}(\rho)}\nonumber\\
E_{(3)}(\rho)&=&E_{(123)}(\rho)+\dots+E_{(n-2,n-1,n)}(\rho)\nonumber\\
\dots&=&\dots\nonumber\\
 E_{(n)}(\rho)&=&E_{(12\dots n)}(\rho)\;.
\end{eqnarray}
We find that for the states under investigation the only
non--vanishing entanglement is  the $n$--partite entanglement and it
derives to (for details next Sec.~\ref{derivationmeasure})
\beq E_{(n)}&=&E_{12\dots
n}\;=\;X\max\big[0,\;\frac{1}{2}\max\biggl[-1+\vec{c}\cdot\vec{n}^{\,(1)},\nonumber\\&&
-1+\vec{c}\cdot\vec{n}^{\,(2)},
-1+\vec{c}\cdot\vec{n}^{\,(3)},-1+\vec{c}\cdot\vec{n}^{\,(4)}\,]\biggr]^2\;,\nonumber\\
\eeq where $X=1$ except for bipartite qubits then it is $X=2$ (the
reason of this difference is explained later). Hence, we find the
same condition for being entangled as given by the one qubit Peres
criterion.

Now, if these bounds are exact also for $n>2$, then all states
represented by the octahedron are separable. Indeed, it turns out
that this is the case. We give the proof of separability separately
in the appendix.

In summary, we have found for even number of qubits the same
geometry as in the case of bipartite qubits, also depicted by
Fig.~\ref{qubitstetrahedron}. Moreover, we have shown that the
multipartite entanglement measure proposed by Ref.~\cite{HH2} works
tightly as the bounds are exact and it reveals only $n$--partite
entanglement. Let us discuss this result more carefully.

For the purest states, $|\vec{c}\,|^2=3$, located in the vertices of
the tetrahedron, the maximal $n$--partite entanglement derives to
$E_{(n)}=1$ except for $n=2$ it is $E_{(n)}=2$. Thus the amount of
entanglement for $n>2$ is independent of the number of qubits
involved. The reason for the difference can be found in the
information content of a multipartite system, Eq.~(\ref{voll}). The
maximal entanglement of a $n$--partite state is $n$. This is the
case if and only if the local obtainable information of all
subsystems is zero and the classical lack of knowledge of the
quantum state is also zero, i.e. the total state is pure. For
bipartite qubits, $n=2$, the vertex states are the Bell states,
which have maximal entanglement $2$ whereas there locally obtainable
information $S$ is zero as well as the lack of classical knowledge
about the quantum state $R=0$.

By construction for $n>2$ we set the locally obtainable information
$S$ of all subsystems zero, however, also all possible locally
obtainable information shared by two, three, \dots, $n-1$ parties is
set to zero; obviously this is not compatible with being maximally
entangled. The information content for $n>2$ is given by \beq n&=&
E_n+R\;=\;1+R\;,\eeq and consequently the lack of classical
knowledge is nonzero, i.e. $R=n-1$. Differently stated for $n=4$,
any party has the trace state as well as any two parties and any
three parties share the trace state, therefore $R=3$.

\noindent\textit{Remark:} The local information $S_s(\rho)$ of one
subsystem $s$ is nothing else than Bohr's quantified complementarity
relation \cite{HH1,SBGH3,HV}, with its well known physical
interpretation in terms of predictability and visibility
(coherence). One can extend this concept for two parties sharing a
state, then their (bi--)local information of total multipartite
system can be defined in similar way and is complemented by the
mixedness of the shared bipartite system. Again this (bi--)local
information is only obtainable if and only if the state is not the
trace state.

Coming back to the simplex geometry we see that the closer we get to
the origin the more the amount of entanglement reduces by increasing
the amount of classical uncertainty $R$ only.

For bipartite qubits the vertex states $|\vec{c}|^2=3$ are the four
Bell states. For $n$ qubits we find for $|\vec{c}|^2=3$ also four
unitary equivalent states, however, they are no longer pure. For
$n=4$ the state is a equally weighted mixture of four $|GHZ\rangle$
states: Starting with one GHZ--state, e.g.
\beq|GHZ\rangle&=&\frac{1}{\sqrt{2}}\big\lbrace|0000\rangle+|1111\rangle\big\rbrace\eeq
one obtains another representation by applying two flips, i.e.
$\mathbbm{1}\otimes\mathbbm{1}\otimes\sigma_x\otimes\sigma_x$, then
applying on the new GHZ--state representation the operator
$\mathbbm{1}\otimes\sigma_x\otimes\sigma_x\otimes\mathbbm{1}$ and
onto that new  GHZ--state representation the operator
$\sigma_x\otimes\sigma_x\otimes\mathbbm{1}\otimes\mathbbm{1}$ gives
the last GHZ--state representation. The other three vertex states
are obtained by applying only one Pauli matrix. For $n=6$ we have
$2^6$ GHZ--states where $2^6/4$ GHZ--states equally mix for one
vertex state.

\noindent\textit{Remark:} The same symmetry we find for the
bipartite qubit case, one Bell states is mapped into another by one
Pauli matrix, however, applying two Pauli matrices maps a Bell state
onto itself, therefore we have no mixture of different maximally
entangled states.

In the next section we give the detailed calculation of the measure
and in the following section we investigate the question whether the
entangled states are bound entangled and if in what sense their
entanglement is bound. In particular we discuss what it means that
the substructure revealed by the measure shows only $n$--partite
entanglement.

\section{Derivation of the multipartite
measure for the simplex states}\label{derivationmeasure}

In Ref.~\cite{HH2} a multipartite measure for multidimensional
systems as a kind of generalization of Bohr's complementarity
relation was derived. Here, we give explicitly the results for $n=2$
and $n=4$ expressed in the familiar Pauli matrix representation

It is well known that to compute concurrence introduced by Hill and
Wootters \cite{Wootters} one  has to consider
\beq\rho\;(\sigma_y\otimes\sigma_y)\;\rho^*\;(\sigma_y\otimes\sigma_y)\eeq
where the complex conjugation is taken in computational basis. The
concurrence is then given by the formula
\beq\mathcal{C}&=&\max\bigl\{0, 2
\max\{\lambda_1,\lambda_2,\lambda_3,\lambda_4\}-(\lambda_1+\lambda_2+\lambda_3+\lambda_4)\bigr\}\nonumber\\\eeq
where the $\lambda_i$'s are the square roots of the eigenvalues of
the above matrix. To obtain the information content we have to
multiply this measure by two.

The first observation in Ref.~\cite{HH2} is that the linear entropy,
$M(\rho)=\frac{2}{3}\left(1-\, Tr(\rho^2)\right)$, can be rewritten
by operators. This means e.g. for any pure $4$ qubit state
\beq|\psi\rangle&=&\sum_{i,j,k,l=0}^1\,
a_{ijkl}\;|ijkl\rangle\;,\eeq the linear entropy of one subsystem
can be written as \begin{widetext}
\begin{eqnarray}\label{psics}
\lefteqn{M^2(Tr_{234}|\psi\rangle\langle\psi|)=M^2(\rho_1)=}\nonumber\\
&&\sum_{k,l=0}^1 \sum_{\{i_1\not= i_1'\};\,\{i_2\not=
i_2'\}}\bigl|\langle\psi|(\sigma_x\otimes\sigma_x\otimes\mathbbm{1}\otimes\mathbbm{1})
\left(|i_1\, i_2\, k\, l\rangle\langle i_1\, i_2\, k\, l|-|i_1'\,
i_2'\, k\, l\rangle\langle
i_1'\, i_2'\, k\, l|\right)|\psi^*\rangle\bigr|^2\nonumber\\
 &+&\sum_{k,l=0}^1
\sum_{\{i_1\not=i_1'\};\,\{i_3\not=i_3'\}}\bigl|\langle\psi|(\sigma_x\otimes\mathbbm{1}\otimes\sigma_x\otimes\mathbbm{1})
\left(|i_1\, k\, i_3\,l\rangle\langle i_1\, k\, i_3\,l|-|i_1'\, k
\,i_3'\,l\rangle\langle i_1'\, k\, i_3'\,l|\right)|\psi^*\rangle\bigr|^2\nonumber\\
 &+&\sum_{k,l=0}^1
\sum_{\{i_1\not=i_1'\};\,\{i_3\not=i_3'\}}\bigl|\langle\psi|(\sigma_x\otimes\mathbbm{1}\otimes\mathbbm{1}\otimes\sigma_x)
\left(|i_1\, k\,l\, i_4\rangle\langle i_1\, k\,l\, i_4|-|i_1'\, k\,l
\,i_3'\rangle\langle i_1'\, k\,l\, i_4'|\right)|\psi^*\rangle\bigr|^2\nonumber\\
 &+&\sum_{k,l=0}^1
\sum_{\{i_2\not=i_2'\};\,\{i_3\not=i_3'\}}\bigl|\langle\psi|(\mathbbm{1}\otimes\sigma_x\otimes\sigma_x\otimes\mathbbm{1})
\left(|k\,i_2\,i_3\, l\rangle\langle k\,i_2\,i_3\,l|-|k\,i_2'\,i_3'\,l\rangle\langle k\,i_2'\,i_3'\,l|\right)|\psi^*\rangle\bigr|^2\nonumber\\
 &+&\sum_{k,l=0}^1
\sum_{\{i_2\not=i_2'\};\,\{i_4\not=i_4'\}}\bigl|\langle\psi|(\mathbbm{1}\otimes\sigma_x\otimes\mathbbm{1}\otimes\sigma_x)
\left(|k\,i_2\, l\,i_4\rangle\langle k\,i_2\, l\,i_4|-|k\,i_2'\, l\,i_4'\rangle\langle k\,i_2'\, l\,i_4'|\right)|\psi^*\rangle\bigr|^2\nonumber\\
 &+&\sum_{k,l=0}^1
\sum_{\{i_3\not=i_3'\};\,\{i_4\not=i_4'\}}\bigl|\langle\psi|(\mathbbm{1}\otimes\mathbbm{1}\otimes\sigma_x\otimes\sigma_x)
\left(|k\,l\,i_3\,i_4\rangle\langle k\,l\,i_3\,i_4|-|k\,l\,i_3'\,i_4'\rangle\langle k\,l\,i_3'\,i_4'|\right)|\psi^*\rangle\bigr|^2\nonumber\\
 &+&\sum_{k}^1 \sum_{\{i_1\not=i_1'\};\,\{i_2\not=i_2'\};\,\{i_3\not=i_3'\}}
\bigl|\langle\psi|(\sigma_x\otimes\sigma_x\otimes\sigma_x\otimes\mathbbm{1})
\left(|i_1\, i_2\, i_3\,k\rangle\langle i_1\, i_2\, i_3\, k|-|i_1'\,
i_2' \,i_3'\,k\rangle\langle i_1'\,
i_2'\,i_3'\,k|\right)|\psi^*\rangle\bigr|^2\nonumber\\
 &+&\sum_{k=0}^1
\sum_{\{i_1\not=i_1'\};\,\{i_2\not=i_2'\};\,\{i_4\not=i_4'\}}
\bigl|\langle\psi|(\sigma_x\otimes\sigma_x\otimes\mathbbm{1}\otimes\sigma_x)
\left(|i_1\, i_2\,k\,i_4\rangle\langle i_1\, i_2\,k\,i_4|-|i_1'\,
i_2'\,k\,i_4'\rangle\langle i_1\,
i_2'\,k\,i_4'|\right)|\psi^*\rangle\bigr|^2\nonumber\\
&+&\sum_{k=0}^1
\sum_{\{i_1\not=i_1'\};\,\{i_3\not=i_3'\};\,\{i_4\not=i_4'\}}
\bigl|\langle\psi|(\sigma_x\otimes\mathbbm{1}\otimes\sigma_x\otimes\sigma_x)
\left(|i_1\,k\, i_3\,i_4\rangle\langle i_1\,k\, i_3\,i_4|-|i_1'\,k\,
i_3'\,i_4'\rangle\langle i_1'\,k\,
i_3'\,i_4'|\right)|\psi^*\rangle\bigr|^2\nonumber\\
&+&\sum_{k=0}^1
\sum_{\{i_2\not=i_2'\};\,\{i_3\not=i_3'\};\,\{i_4\not=i_4'\}}
\bigl|\langle\psi|(\mathbbm{1}\otimes\sigma_x\otimes\sigma_x\otimes\sigma_x)
\left(|k\,i_2\, i_3\,i_4\rangle\langle k\,i_2\, i_3\,i_4|-|k\,i_2'\,
i_3'\,i_4'\rangle\langle k\,i_2'\,
i_3'\,i_4'|\right)|\psi^*\rangle\bigr|^2\nonumber\\
&+&\sum_{\{i_1\not=i_1'\};\,\{i_2\not=i_2'\};\,\{i_3\not=i_3'\};\,\{i_4\not=i_4'\}}
\bigl|\langle\psi|(\sigma_x\otimes\sigma_x\otimes\sigma_x\otimes\sigma_x)
\left(|i_1\,i_2\, i_3\,i_4\rangle\langle i_1 \,i_2\,
i_3\,i_4|-|i_1'\,i_2'\, i_3'\,i_4'\rangle\langle i_1'\,i_2'\,
i_3'\,i_4'|\right)|\psi^*\rangle\bigr|^2
\end{eqnarray}
where e.g. $\{i_1\}\not=\{i_1'\},\{i_2\}\not=\{i_2'\}$ means that
the set of indexes are not the same, i.e. the sum is taken over
\beq\{i_1,i_2;i_1',i_2'\}&=&\{0,1;0,0\},\{0,0;0,1\},\{0,1;1,0\},\{0,0;1,1\},\{1,1;0,0\},\{1,0;0,1\},\{1,1;0,0\},\{1,0;0,1\},\nonumber\\
&&\{0,0;1,0\},\{1,0;0,0\},\{0,0;1,1\},\{1,0;0,1\},\{0,1;1,0\},\{1,1;0,0\},\{0,1;1,1\},\{1,1;0,1\}\;.\eeq
\end{widetext}
Likewise the linear entropies for the other subsystem can be
derived, i.e. separated in terms where the flip operator $\sigma_x$
is applied two, three or four times. It is well known that for pure
states the sum over the entropies of all reduced density matrices is
an entanglement measure, therefore using the linear entropy we get
the following entanglement measure \beq
E(|\psi\rangle):=\sum_{s=1}^4 M^2(\rho_s)\;=\; \sum_{m=2}^{4}
\left(C^{m}(\psi)\right)^2\,,\eeq where $(C^{m})^2$ is the sum of
all terms of all reduced matrices which contain $m$--flip operators.
These quantities where called (squared) $m$--concurrences, because
they play a similar role as Wootters concurrence.

For mixed states $\rho$ the infimum of all possible decompositions
is an entanglement measure \beq
E(\rho)\;=\;\inf_{p_i,|\psi_i\rangle}\sum_{p_i,|\psi_i\rangle} p_i
E(|\psi_i\rangle)\;.\eeq The problem of the whole entanglement
theory is that this infimum can in general not be calculated. Now we
bring the operator representation of the linear entropy into the
game, because for operators upper bounds can be obtained.

Lets start with the calculation of the $4$-flip concurrence
$C^{(4)}$, which is the sum of all terms containing $4$-flips of the
entropies of all reduced matrices, i.e. \beq
\left(C^{(4)}(\rho)\right)^2&=&\inf_{p_i,|\psi_i\rangle}\;\sum_{p_i,|\psi_i\rangle}
p_i \left(C^{(4)}(\psi_i)\right)^2
\eeq As shown in Ref.~\cite{HH2} one can derive bounds on the above
expression for any $m$--flip concurrence by defining, in an
analogous way to Hill and Wootters flip density matrix
\cite{Wootters}, the $m$--flip density matrix:
\begin{eqnarray}
\widetilde{\rho}_{s}^m=O_{s}(|\{i_n\}\rangle
\langle\{i_n\}|-|\{i'_n\}\rangle\langle\{i'_n\}|)\;\rho^*
\cdot\nonumber\\ \cdot\; O_{s}
(|\{i_n\}\rangle\langle\{i_n\}|-|\{i'_n\}\rangle\langle\{i'_n\}|)
\end{eqnarray}
and calculating the $\lambda_m^{s}$'s which are the squared roots of
the eigenvalues of $\widetilde{\rho}_{s}^m \rho$. The bound
$B^{(m)}$ of the $m$-flip concurrence $C^{(m)}$ is then given by
\begin{eqnarray}\label{bounds}
\lefteqn{B^m(\rho):=}\nonumber\\
&&\left(\sum_{s}\textbf{max}\left[0,2\textbf{max}
\left(\{\lambda_m^{s}\}\right)-\sum
\{\lambda_m^{s}\}\right]^2\right)^{\frac{1}{2}}\nonumber\\
\end{eqnarray}

From Eq.~(\ref{psics}) we see that for the $4$--flip concurrence of
subsystem $\rho_1$ four different operators occur, thus we have in
total $16$ different operators listed in the appendix
\ref{appendix1}.

Inserting our class of states we find that for each operator
$\mathcal{O}^{s}$ the eigenvalues are the same, i.e. one obtains $8$
zeros and the remaining four eigenvalues are exactly equivalent to
the Peres criterion Eq.~(\ref{PTsingle}).

The same procedure has to be applied to calculate the $3$--flip
concurrence and the $2$--flip concurrence. As can be seen from
Eq.~(\ref{psics}) here the unity and $\sigma_z$ matrix is involved
which lead to no contribution for the states under investigation.
Remember, that they are mixtures of the vertex states, which are
equal mixtures of such GHZ--states which differ by two flips.

Therefore, the total entanglement is given by the $C^{(4)}$
concurrence only and is a $4$--partite entanglement. For
$n=6,8,\dots$ the scenario is the same, because of the same
underlying symmetry.

In the Appendix we show that all states not detected by the measure
are separable, thus the bounds are optimal and therefore
the measure detects all bound entangled states. 
\section{Are the entangled states bound entangled?}\label{bound}

In
Refs.~\cite{Smolin,HorodeckigeneralizedSmolin2,HorodeckigeneralizedSmolin,Bandyopadhyay04}
the special states $c=c_1=-c_2=c_3$ for $n>2$, which were named
generalized Smolin states (for $n=2$ these states are the Werner
states), are investigated and they show that for $1\geq
c>\frac{1}{3}$ these states are bound entangled. In particular, the
authors argued that these states are bound entangled, because the
states are separable against bipartite symmetric cuts like
$12|34\dots, 14|23\dots,\dots$ and therefore no Bell state between
any two subsystem can be distilled. This is obviously also the case
for the whole class of states under investigation.

As the considered measure of entanglement revealed only $n$--partite
entanglement and e.g. not any $m$--partite entanglement ($m<n$), it
may not seem directly obvious that Bell states (bipartite
entanglement) cannot be distilled, because the class of states do
not possess any bipartite entanglement. Thus the question could be
refined to ask whether $n$--partite pure entanglement can be
distilled.

\begin{figure}
\center{\includegraphics[height=100pt,
keepaspectratio=true]{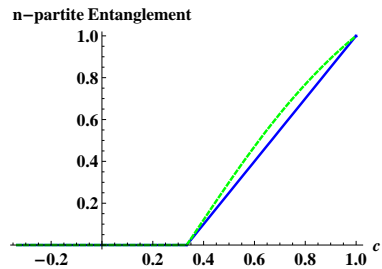}} \caption{(Color online) Here
the $n$--partite entanglement of the Werner states $n=2$ (here the
$y$--axis has to be multiplied by two) or the generalized Smolin
states $n>2$ before and after the application of the introduced
protocol (upper dashed green curve) is plotted. Note that the vertex
states are mapped onto itself by the given
protocol.}\label{diskurvefig}
\end{figure}

\begin{figure*}
\center{ (a)\includegraphics[height=170pt,
keepaspectratio=true]{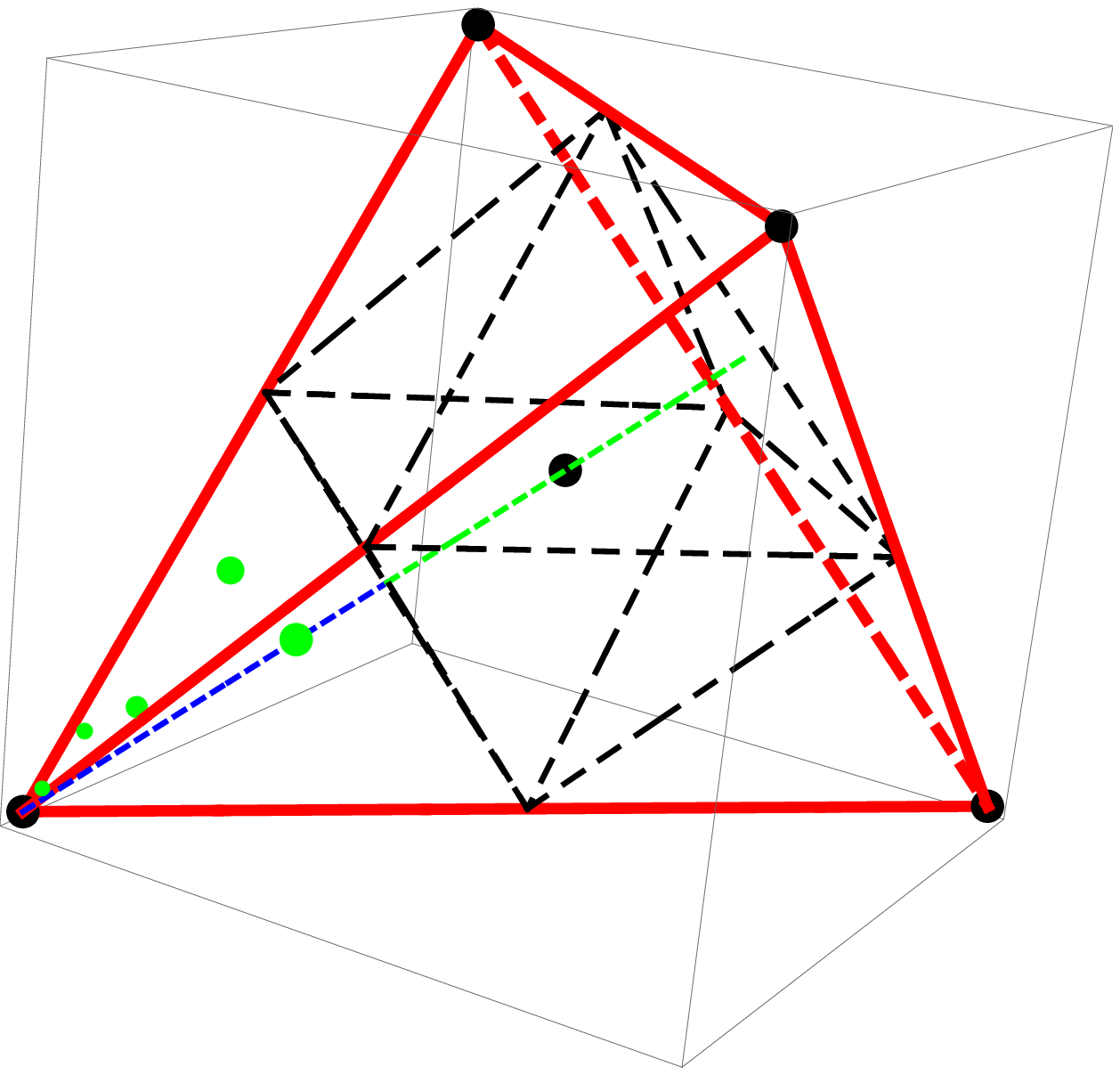}\hspace{0.5cm}
(b)\includegraphics[height=150pt,
keepaspectratio=true]{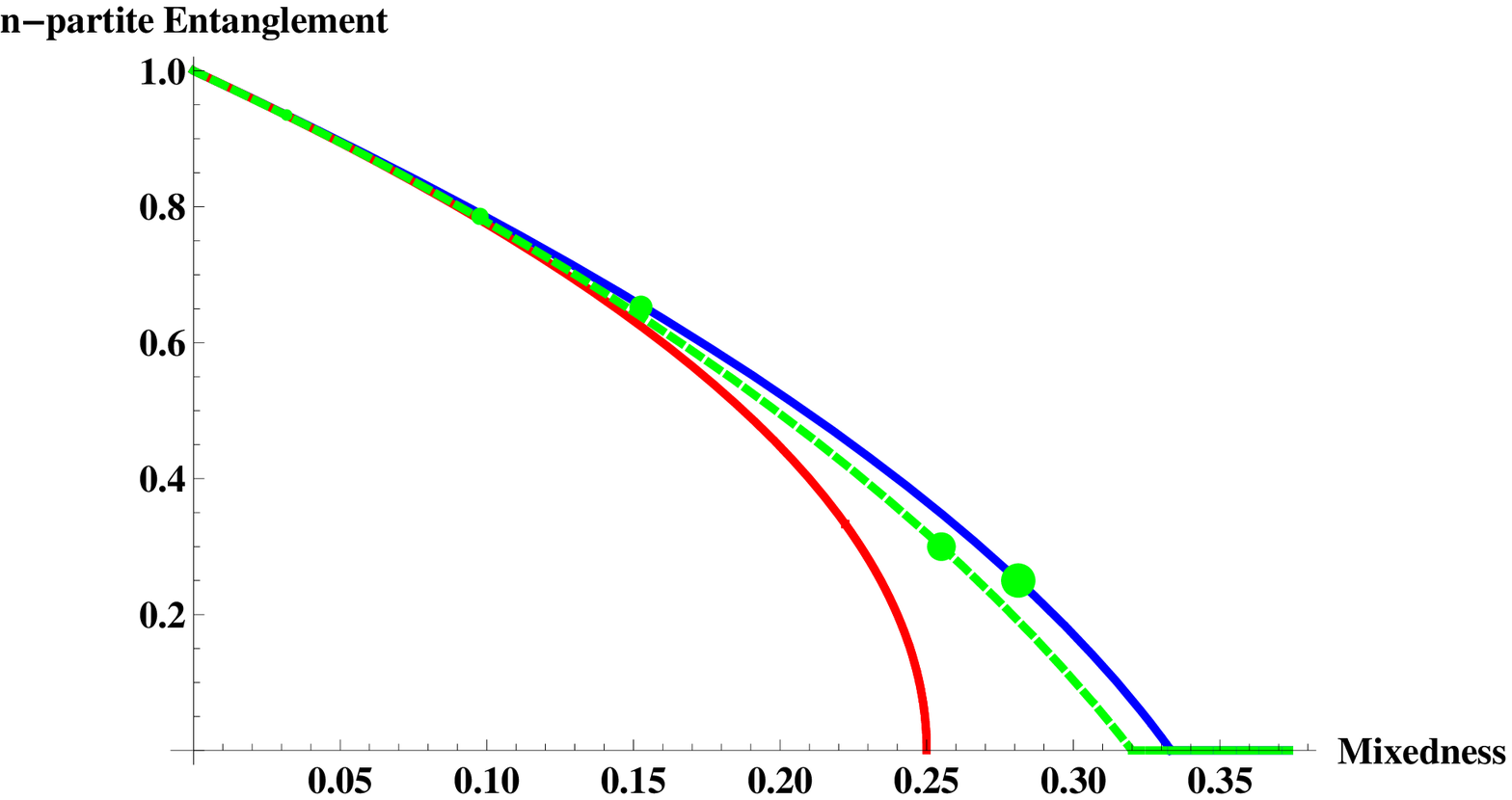}} \caption{(Color online)
Fig.~(a) shows the final states after each step of the introduced
protocol of an initial Werner or Smolin state $c=0.5$, where each
(green) point represents the obtained state after one step of the
protocol. Fig.~(b) shows a mixedness,
$\frac{2^n}{2^n-1}(1-Tr(\rho^2))$, versus $n$--partite entanglement
diagram (for $n=2$ the $y$--axis has to be multiplied by $2$), where
the (blue) curve corresponds to the Werner or Smolin state whereas
the (red) curve is the state connecting two vertices. All states of
the simplex have their mixedness--entanglement ratio between these
two curves. The middle (dashed, green) curve corresponds to the
final states of a distilled Werner or Smolin state. And the (green)
points represent the final states after each step of an initial
Werner or Smolin state $c=0.5$.}\label{disfig}
\end{figure*}


For the $n$--partite class of states under investigation we consider
a similar distillation protocol as the recurrence protocol by
Bennett et al. \cite{Bennettdis}. For that we generalize it such
that each party gets a copy onto which a unitary bilateral XOR
operation is performed and afterwards a measurement in say
$z$-direction is performed. Only states are kept where all parties
found their copy qubit in say up--direction. This protocol favours
as all protocols do one state, in our case for $n=2$ it is the
$\Phi^+$ state and for $n>2$ its equivalents.

In detail it goes like the following: We consider one state and its
copy \beq \rho^{\otimes 2}&=&\left(\frac{1}{2^n}\left\lbrace
\mathbbm{1}^{\otimes n}+c_i\; \sigma_i^{\otimes
n}\right\rbrace\right)^{\otimes 2}\eeq and all parties get a copy
state. Therefore, we reorder the state by a unitary transformation
such that the first term and second term in the tensor product
belongs to Alice and the third and forth term to Bob and so on: \beq
\rho^{\otimes 2}&\longrightarrow&(\frac{1}{2^n})^2\biggl\lbrace
(\mathbbm{1}\otimes\mathbbm{1})^{\otimes n}+c_i\;(\mathbbm{1}\otimes
\sigma_i)^{\otimes
n}\nonumber\\
&&+c_i\; (\sigma_i\otimes \mathbbm{1})^{\otimes n}+\;c_i c_j\;
(\sigma_i\otimes \sigma_j)^{\otimes n}\biggr\rbrace\eeq Now each
party perform on its two subsystems a unitary XOR operation \beq
U_{XOR}&=&\left(\begin{array}{cccc} 1&0&0&0\\
0&1&0&0\\
0&0&0&1\\0&0&1&0\\\end{array}\right)\eeq and then projects on the
copy--subsystem with $P=\frac{1}{2}(\mathbbm{1}+\sigma_z)$. This
gives again a state in the class of states under investigation, i.e.
one finds
 \beq
\vec{c}&=&\left(\begin{array}{c}c_x\\c_y\\c_z\end{array}\right)\qquad\longrightarrow
\vec{c}_{\rm{dis}}=\left(\begin{array}{c}\frac{c_x^2+c_y^2}{1+c_z^2}\\\frac{2
c_x c_y}{1+c_z^2}\\\frac{2 c_z}{1+c_z^2}\end{array}\right)\;.\eeq
Comparing with the separability condition and with the positivity
condition, one verifies that only separable states are mapped into
separable states.

Let us consider the Werner states and the generalized Smolin states
($c=c_x=c_y=c_z$), for which we derive that the $n$--partite
entanglement is always increased after the above protocol, see
Fig.~\ref{diskurvefig}. For $-\frac{1}{\sqrt{3}}\leq
c\leq\frac{1}{3}$ the measure before and after the protocol is zero
and for $c=1$ the state is mapped onto itself. For $\frac{1}{3}<c<1$
the entanglement of the distilled state is increased compared to the
input state. In Fig.~\ref{disfig}~(a) we give the $3$--dimensional
picture of how the initial state $c=0.5$ moves after each step
towards the vertex state. Note that the states are no longer in the
set of the generalized Smolin sets, another advantage of considered
set of states as no random bilateral rotation to regain the
rotational symmetry is needed. In Fig.~\ref{disfig}~(b) we show the
mixedness--entanglement relation of this example. Note that all
states of the simplex are within the two curves and the middle curve
is the result for the generalized Smolin state after one step of the
protocol.

\noindent\textit{Remark:} Not all states of the simplex are mapped
into more entangled states by this protocol. For example, the
mixture of two vertex states ($\vec{c}^{\,T}=(0,0,c)$ with
$c\not=1$) is left invariant.

In summary, we have found a protocol that increases the amount of
entanglement with local operations and classical communication only
and the final states are always within the class of states. Only for
$n=2$ the final state is pure and maximal entangled and therefore
the above protocol is a distillation protocol, i.e. pure maximally
entangled states can be obtained. However, for $n>2$ the final state
is no longer pure, but has the maximal $n$--partite entanglement of
the class of states under investigation.

Thus the next logical step is to search for a distillation protocol
which distills the vertex states into pure maximally entangled
states, i.e. GHZ--states. However this is not possible for the
following reasons: In general, any equally weighted mixture of two
maximally entangled states cannot be distilled by mainly two
observations. As for all maximally entangled states $\rho_i$
obviously the entanglement can only be reduced by any completely
positive map $\Lambda: \rho_i\mapsto\rho_i'$, i.e. $E(\rho_i')\leq
E(\rho_i)\; \forall\; \Lambda$. And as the entanglement $E(\rho)$ is
convex, i.e. $E(\rho_i')+E(\rho_j')\leq 2 E(\rho_i')$, we conclude
that at least one $\rho_i$ must be mapped unitary onto itself or
another maximally entangled state. Because all maximally entangled
states are equivalent by local unitaries, such a map consequently
maps also the other maximally entangled state of the mixture into a
(different) maximally entangled state. Hence, for no equally mixture
of maximally entangled states a maximally entangled state can be
distilled. Note that in the case of bipartite qubits this is
trivially true, because any equally mixture of Bell states is
separable, however, for multipartite states this is not necessarily
the case (e.g. our vertex states).

Thus we find that we can increase the amount of the $n$--partite
entanglement until the vertex state, but not furthermore and
therefore all entangled states are bound entangled, i.e. no pure
$n$--partite entanglement can be distilled among any subset of
parties using stochastic LOCC. The common definition of distillation
is that no pure maximally entanglement among any subset of parties
using LOCC can be obtained, see e.g.
\cite{DurCirac,ShorSmolinThapliyal}. A different way to prove that
the entangled states are bound is given in
Ref.~\cite{HorodeckiBipartGHZ}, where they show that if no singlets
can be distilled also no GHZ---state can be obtained. Therefore for
the class of states under investigation we can also not distill any
bipartite entanglement.

\section{The geometry of the states violating the CHSH---Bell
inequality}\label{chshsimplex}

Analog to the bipartite qubit state one can derive a CHSH--Bell type
inequality for $n$ qubit states \cite{CHSHmulti}. Here $n-1$ parties
measure their qubit in direction $\vec{a}$ or $\vec{a}'$ and the
$n$th party in direction $\vec{b}$ or $\vec{b}'$, then one obtains
the following Bell inequality \beq\label{Belline}
Tr(\mathcal{B}_{\textrm{Bell-CHSH}}\rho)\;\leq\; 2\eeq with \beq
\mathcal{B}_{\textrm{Bell-CHSH}}&=&
\underbrace{\vec{a}\vec{\sigma}\otimes
\vec{a}\vec{\sigma}\otimes\dots\otimes
\vec{a}\vec{\sigma}}_{n-1}\otimes (\vec{b}+\vec{b}')
\vec{\sigma}\nonumber\\
&+& \underbrace{\vec{a}'\vec{\sigma}\otimes
\vec{a}'\vec{\sigma}\otimes\dots\otimes
\vec{a}'\vec{\sigma}}_{n-1}\otimes (\vec{b}-\vec{b}')
\vec{\sigma}\;\eeq where $\vec{a},\vec{a}',\vec{b},\vec{b'}$ are
real unit vectors and the value $2$ is the upper bound on any local
realistic theory.

It is known that for $n=2$ the maximal violation by quantum
mechanics can simply be derived by the state $\rho$ itself
\cite{HorodeckiCHSH}. A matrix $\rho$ violates the Bell--CHSH
inequality if and only if ${\cal M}(\rho)\geq 1$, where ${\cal
M}(\rho)$ is the sum of the two largest eigenvalues of the Hermitian
matrix $C^\dagger C$ with $(C)_{ij}=Tr(\sigma_i\otimes\sigma_j
\rho)$. A generalization for $n$ qubits is simple, because the
matrix $C$ is diagonal for the states under investigation, thus the
same proof works.

\begin{figure}
\center{
\includegraphics[width=200pt, keepaspectratio=true]{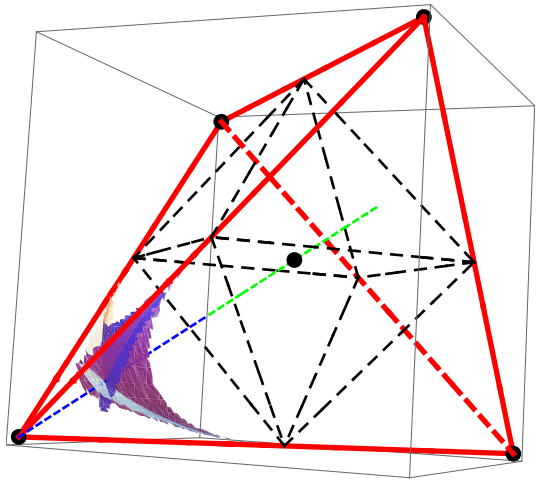}}
\caption{(Color online) The three cylinders show the saturation of
the Bell inequality. All states outside these cylinders violate the
Bell inequality. The vertex states violate the Bell inequality
maximal, i.e. by $2\sqrt{2}$.}\label{chshfig}
\end{figure}

In our case ${\cal M}(\rho)$ is simply the sum of the two largest
squared vector components. In particular, if $c_1$ and $c_2$ are
greater than $c_3$ we obtain the following Bell inequality \beq
c_1^2+c_2^2\leq 1\;.\eeq This gives a simple geometric
interpretation of all states violating the Bell inequality. All
possible saturated Bell inequalities give three different cylinders
in the picture representing the state space, see Fig.\ref{chshfig}.
All states outside of these three cylinders violate the Bell
inequality.

Furthermore, this result shows that an entangled state not violating
the Bell inequality (\ref{Belline}), can be transformed via the
introduced protocol into a state violating the Bell inequality,
leading to the conclusion that all entangled states of the picture
have nonlocal features. Moreover, in agreement with Ref.~\cite{Dur}
the possibility to construct realistic local models or not is no
criterion for being bound entangled or not.

Let us also remark that Werner states ($n=2$) violate the Bell
inequality for $c>\frac{1}{\sqrt{2}}$ whereas successful
teleportation requires only $c>\frac{1}{2}$.

\section{Summary and Discussion}

We generalized the magic simplex for locally maximally mixed
bipartite qubit states such that we add even numbers $n$ of qubits
and set all partial traces equal to the maximally mixed states, i.e.
no local information obtainable by any subset of parties is
available. This class of states can be described by three real
numbers which enables us to draw a three dimensional picture.
Interestingly, we find the same geometry concerning separability,
mixedness, kind of entanglement, amount of entanglement and
nonlocality for all even numbers of qubits, see also
Fig.~\ref{qubitstetrahedron} and Fig.~\ref{chshfig}.

For $n>2$ the purest states, located in the vertices of the simplex,
are not pure except in the case of bipartite qubits ($n=2$). We show
how to derive a recently proposed measure for all discrete
multipartite systems \cite{HH2} in this case. For mixed states only
bounds exist, however, we show that they are for the class of states
optimal by proving that all states not detected by the measure are
separable.

The measure reveals that these states only possess $n$--partite
entanglement and no other kind of entanglement, e.g. bipartite
entanglement. The information content of the states can be
quantified by the generalized Bohr's complementarity relation for
$n>2$ \beq n&=&\mathcal{S}+E_n+R\;=\;1+R\;,\eeq where $R$ lack of
classical knowledge and $\mathcal{S}=0$ the local information
obtainable by any party.

Then we investigated the question whether the $n$--partite
entanglement can be distilled. We find a protocol using only local
operation and classical communication (LOCC) which increases the
$n$--partite entanglement to the maximal entanglement of the class
of states under investigation. These states are the vertex states of
the simplex, for $n=2$ they are the Bell states and for $n>2$ they
are equal mixtures of such GHZ--states which are obtained by
applying only two flips, $\sigma_x$.

For bipartite qubits $n=2$ this protocol is a distillation protocol,
i.e. pure maximally entangled states are obtained. For $n>2$ the
vertex states are not pure, therefore we search for a distillation
protocol that leaves the class of states under investigation to
obtain a pure $n$--partite maximally entangled state, i.e. the
GHZ--states. Indeed, we argue that such a protocol cannot be found,
more precisely, any equal mixture of GHZ--states cannot be
distilled. Thus for the class of states under investigation all
entangled states are bound entangled and herewith we found a simplex
where all states are either separable or bound entangled.

In detail, we show how an initial state moves after each step of the
protocol increasing the entanglement in the simplex, see
Fig.~\ref{diskurvefig}. Moreover, we find that the states violating
the CHSH--Bell like inequality, which was shown to be optimal in
this case, have for all even numbers of qubits the same geometry,
see Fig.~\ref{chshfig}. These two results taken together mean that
one can enhance the $n$--partite bound entanglement by only using
LOCC until the Bell inequality is violated. Therefore, for all
$n$--partite bound entangled states its (hidden) nonlocality is
revealed and in agreement with Ref.~\cite{Dur} a possibility whether
a local realistic theory can be constructed is not a criterion for
distillability and likewise whether its entanglement can be
increased by LOCC is also no criterion.

%

Our results suggest also that one can distinguish between bound
states for which a certain entanglement measure cannot be increased
by LOCC (in our case the vertex states) and states for which the
entanglement can be increased by LOCC, which may be denoted by
``quasi'' bound entangled states (all bound entangled states of the
class except the vertex states). The introduced (distillation)
protocol distills maximally entangled states within the set of
states which are, however, not pure, but the purest of the set of
states.

Last but not least we want to remark that a subset of the class of
states was considered in literature, e.g.
\cite{Smolin,HorodeckigeneralizedSmolin2,HorodeckigeneralizedSmolin,Bandyopadhyay04},
the so called Smolin states. For which it was shown that no Bell
states may be distilled. The theorem in
Ref.~\cite{HorodeckiBipartGHZ} states that if and only if bipartite
entanglement can be distilled then also $GHZ$-states ---in our
terminology $n$--partite entanglement--- can be distilled.

In summary, we have shown in this paper explicitly that the
multipartite measure proposed by \cite{HH2} detects all bound
entanglement in the class of states and that the states do not
possess bipartite entanglement and how the $n$--partite entanglement
can be increased to a certain value.


%
%

These results do not only help to reveal the mysteries of bound
entanglement by refining its kind of entanglement, but they may also
help to construct quantum communication scenarios where bound
entangled states actually help to perform a certain process
\cite{Horodeckiactivatingbound}. This is clearly important, when one
has future application in mind, e.g. a multipartite cryptography
scenario.

\textbf{Acknowledgement:} Many thanks to B. Baumgartner, R.A.
Bertlmann, W.~D\"ur and R.~Augusiak for enlightening discussions.


\appendix{\noindent\textbf{Appendix: Proof that all states represented by the octahedron are
separable.}}\label{appendix2}

To prove that all states represented by the octahedron are
separable, we show that this is the case for the following points in
the octahedron
\beq\vec{c}=\left(\begin{array}{c}1\\0\\0\end{array}\right),
\left(\begin{array}{c}0\\1\\0\end{array}\right),
\left(\begin{array}{c}0\\0\\1\\\end{array}\right)\;.\eeq As any
convex combination of separable states have to be also separable, we
have finalized the proof. We start with $n=2$ and show how this
construction generalizes for $n=4,6,\dots$.

Suppose Alice prepares her qubits in the following two states: \beq
\omega^A_{i,\pm}&=&\frac{1}{\sqrt{2}}(\mathbbm{1}_2\pm
r^A_i\sigma_i)\;,\eeq where $r_i$ is a Bloch vector pointing in
$i$--direction and is given by any number in $[-1,1]$. Bob does
prepares his qubits in the very same way. If Alices chooses the
positive $i$--axis and Bob does the same, if Alice chooses the
negative sign, Bob does the same, thus they share the following
separable state if the preparation is done randomly with the same
probability: \beq
\rho^{AB}_{i,+}&=&\frac{1}{2}\;\omega^A_{i,^+}\otimes\omega^B_{i,+}
+\frac{1}{2}\;\omega^A_{i,^-}\otimes\omega^B_{i,-}\nonumber\\
&=&\frac{1}{4}(\mathbbm{1}_4+r^A_i\cdot
r^B_i\;\sigma_i\otimes\sigma_i)\;.\eeq These states represent three
vertices of the octahedron, thus the proof is finalized for $n=2$.

Explicitly, we find that for the generalized Smolin state
($c_1=c_2=c_3=c$), the following state derives \beq\rho_c&=&
\sum_i\frac{1}{3}\;
\rho^{AB}_{i,+}\;=\;\frac{1}{4}(\mathbbm{1}_4+\sum_i\frac{r^A_i\cdot
r^B_i}{3}\;\sigma_i\otimes\sigma_i)\;,\eeq therefore as $r^A_i\cdot
r^B_i\in[-1,1]$ the generalized Smolin state is separable for
$p\in[-\frac{1}{3},\frac{1}{3}]$.

For $n=4$ we remark that with the combination \beq
\rho^{AB}_{i,-}&=&\frac{1}{2}\omega^A_{i,^+}\otimes\omega^B_{i,-}
+\frac{1}{2}\omega^A_{i,^-}\otimes\omega^B_{i,+}\nonumber\\
&=&\frac{1}{4}(\mathbbm{1}_4-r^A_i\cdot
r^B_i\;\sigma_i\otimes\sigma_i)\eeq one obtains the minus sign, and
for the very same construction Alice, Bob, Charly and Daisy obtain
the following separable states \beq
\rho^{AB}_{i,+}&=&\frac{1}{2}\;\rho^{AB}_{i,^+}\otimes\rho^{CD}_{i,+}
+\frac{1}{2}\;\rho^{AB}_{i,^-}\otimes\rho^{CD}_{i,-}\nonumber\\
&=&\frac{1}{4}(\mathbbm{1}_4+r^A_i\cdot r^B_i\cdot r^C_i\cdot
r^D_i\;\sigma_i\otimes\sigma_i\otimes\sigma_i\otimes\sigma_i)\;.\eeq
As the combination $+-,-+$ gives again the minus sign this proof
generalizes for any even $n$.

\appendix{\noindent\textbf{Appendix: All $4$--flip operators for $n=4$:}}\label{appendix1}
For convenience of the reader we list all $4$--flip operators in the
Pauli--matrix representation: \beq
\mathcal{O}^1\;=\;\frac{1}{4}\lbrace&&\sigma_y\otimes\sigma_y\otimes\sigma_y\otimes\sigma_y\nonumber\\
&-&\sigma_y\otimes\sigma_y\otimes\sigma_x\otimes\sigma_x\nonumber\\
&-&\sigma_y\otimes\sigma_x\otimes\sigma_y\otimes\sigma_x\nonumber\\
&-&\sigma_y\otimes\sigma_x\otimes\sigma_x\otimes\sigma_y\rbrace\nonumber\\
\mathcal{O}^2\;=\;\frac{1}{4}\lbrace&&\sigma_y\otimes\sigma_y\otimes\sigma_y\otimes\sigma_y\nonumber\\
&-&\sigma_y\otimes\sigma_y\otimes\sigma_x\otimes\sigma_x\nonumber\\
&+&\sigma_y\otimes\sigma_x\otimes\sigma_y\otimes\sigma_x\nonumber\\
&+&\sigma_y\otimes\sigma_x\otimes\sigma_x\otimes\sigma_y\rbrace\nonumber\\
\mathcal{O}^3\;=\;\frac{1}{4}\lbrace&&\sigma_y\otimes\sigma_y\otimes\sigma_y\otimes\sigma_y\nonumber\\
&+&\sigma_y\otimes\sigma_y\otimes\sigma_x\otimes\sigma_x\nonumber\\
&-&\sigma_y\otimes\sigma_x\otimes\sigma_y\otimes\sigma_x\nonumber\\
&+&\sigma_y\otimes\sigma_x\otimes\sigma_x\otimes\sigma_y\rbrace\nonumber\\
\mathcal{O}^4\;=\;\frac{1}{4}\lbrace&&\sigma_y\otimes\sigma_y\otimes\sigma_y\otimes\sigma_y\nonumber\\
&+&\sigma_y\otimes\sigma_y\otimes\sigma_x\otimes\sigma_x\nonumber\\
&+&\sigma_y\otimes\sigma_x\otimes\sigma_y\otimes\sigma_x\nonumber\\
&-&\sigma_y\otimes\sigma_x\otimes\sigma_x\otimes\sigma_y\rbrace \eeq

\beq
\mathcal{O}^5\;=\;\frac{1}{4}\lbrace&&\sigma_y\otimes\sigma_y\otimes\sigma_y\otimes\sigma_y\nonumber\\
&-&\sigma_x\otimes\sigma_y\otimes\sigma_x\otimes\sigma_y\nonumber\\
&-&\sigma_x\otimes\sigma_y\otimes\sigma_y\otimes\sigma_x\nonumber\\
&-&\sigma_y\otimes\sigma_y\otimes\sigma_x\otimes\sigma_x\rbrace\nonumber\\
\mathcal{O}^6\;=\;\frac{1}{4}\lbrace&&\sigma_y\otimes\sigma_y\otimes\sigma_y\otimes\sigma_y\nonumber\\
&-&\sigma_x\otimes\sigma_y\otimes\sigma_x\otimes\sigma_y\nonumber\\
&+&\sigma_x\otimes\sigma_y\otimes\sigma_y\otimes\sigma_x\nonumber\\
&+&\sigma_y\otimes\sigma_y\otimes\sigma_x\otimes\sigma_x\rbrace\nonumber\\
\mathcal{O}^7\;=\;\frac{1}{4}\lbrace&&\sigma_y\otimes\sigma_y\otimes\sigma_y\otimes\sigma_y\nonumber\\
&+&\sigma_x\otimes\sigma_y\otimes\sigma_x\otimes\sigma_y\nonumber\\
&-&\sigma_x\otimes\sigma_y\otimes\sigma_y\otimes\sigma_x\nonumber\\
&+&\sigma_y\otimes\sigma_y\otimes\sigma_x\otimes\sigma_x\rbrace\nonumber\\
\mathcal{O}^8\;=\;\frac{1}{4}\lbrace&&\sigma_y\otimes\sigma_y\otimes\sigma_y\otimes\sigma_y\nonumber\\
&+&\sigma_x\otimes\sigma_y\otimes\sigma_x\otimes\sigma_y\nonumber\\
&+&\sigma_x\otimes\sigma_y\otimes\sigma_y\otimes\sigma_x\nonumber\\
&-&\sigma_y\otimes\sigma_y\otimes\sigma_x\otimes\sigma_x\rbrace \eeq
\beq
\mathcal{O}^9\;=\;\frac{1}{4}\lbrace&&\sigma_y\otimes\sigma_y\otimes\sigma_y\otimes\sigma_y\nonumber\\
&-&\sigma_x\otimes\sigma_x\otimes\sigma_y\otimes\sigma_y\nonumber\\
&-&\sigma_x\otimes\sigma_y\otimes\sigma_y\otimes\sigma_x\nonumber\\
&-&\sigma_y\otimes\sigma_x\otimes\sigma_y\otimes\sigma_x\rbrace\nonumber\\
\mathcal{O}^{10}\;=\;\frac{1}{4}\lbrace&&\sigma_y\otimes\sigma_y\otimes\sigma_y\otimes\sigma_y\nonumber\\
&-&\sigma_x\otimes\sigma_x\otimes\sigma_y\otimes\sigma_y\nonumber\\
&+&\sigma_x\otimes\sigma_y\otimes\sigma_y\otimes\sigma_x\nonumber\\
&+&\sigma_y\otimes\sigma_x\otimes\sigma_y\otimes\sigma_x\rbrace\nonumber\\
\mathcal{O}^{11}\;=\;\frac{1}{4}\lbrace&&\sigma_y\otimes\sigma_y\otimes\sigma_y\otimes\sigma_y\nonumber\\
&+&\sigma_x\otimes\sigma_x\otimes\sigma_y\otimes\sigma_y\nonumber\\
&-&\sigma_x\otimes\sigma_y\otimes\sigma_y\otimes\sigma_x\nonumber\\
&+&\sigma_y\otimes\sigma_x\otimes\sigma_y\otimes\sigma_x\rbrace\nonumber\\
\mathcal{O}^{12}\;=\;\frac{1}{4}\lbrace&&\sigma_y\otimes\sigma_y\otimes\sigma_y\otimes\sigma_y\nonumber\\
&+&\sigma_x\otimes\sigma_x\otimes\sigma_y\otimes\sigma_y\nonumber\\
&+&\sigma_x\otimes\sigma_y\otimes\sigma_y\otimes\sigma_x\nonumber\\
&-&\sigma_y\otimes\sigma_x\otimes\sigma_y\otimes\sigma_x\rbrace \eeq
\beq
\mathcal{O}^{13}\;=\;\frac{1}{4}\lbrace&&\sigma_y\otimes\sigma_y\otimes\sigma_y\otimes\sigma_y\nonumber\\
&-&\sigma_x\otimes\sigma_x\otimes\sigma_y\otimes\sigma_y\nonumber\\
&-&\sigma_x\otimes\sigma_y\otimes\sigma_x\otimes\sigma_y\nonumber\\
&-&\sigma_y\otimes\sigma_x\otimes\sigma_x\otimes\sigma_y\rbrace\nonumber\\
\mathcal{O}^{14}\;=\;\frac{1}{4}\lbrace&&\sigma_y\otimes\sigma_y\otimes\sigma_y\otimes\sigma_y\nonumber\\
&-&\sigma_x\otimes\sigma_x\otimes\sigma_y\otimes\sigma_y\nonumber\\
&+&\sigma_x\otimes\sigma_y\otimes\sigma_x\otimes\sigma_y\nonumber\\
&+&\sigma_y\otimes\sigma_x\otimes\sigma_x\otimes\sigma_y\rbrace\nonumber\\
\mathcal{O}^{15}\;=\;\frac{1}{4}\lbrace&&\sigma_y\otimes\sigma_y\otimes\sigma_y\otimes\sigma_y\nonumber\\
&+&\sigma_x\otimes\sigma_x\otimes\sigma_y\otimes\sigma_y\nonumber\\
&-&\sigma_x\otimes\sigma_y\otimes\sigma_x\otimes\sigma_y\nonumber\\
&+&\sigma_y\otimes\sigma_x\otimes\sigma_x\otimes\sigma_y\rbrace\nonumber\\
\mathcal{O}^{16}\;=\;\frac{1}{4}\lbrace&&\sigma_y\otimes\sigma_y\otimes\sigma_y\otimes\sigma_y\nonumber\\
&+&\sigma_x\otimes\sigma_x\otimes\sigma_y\otimes\sigma_y\nonumber\\
&+&\sigma_x\otimes\sigma_y\otimes\sigma_x\otimes\sigma_y\nonumber\\
&-&\sigma_y\otimes\sigma_x\otimes\sigma_x\otimes\sigma_y\rbrace \eeq

\end{document}